\documentstyle[aps,prl,multicol,epsf]{revtex}

\begin{document}

%\draft

\title{A New Approach to Large-Scale Nuclear Structure Calculations}

\author{J. Dukelsky$^1$ and S.~Pittel$^2$}

\address{ $^1$ Instituto de Estructura de la Materia, Consejo
Superior de Investigaciones Cientificas, Serrano 123, 28006 Madrid,
Spain \\ $^2$Bartol Research Institute, University of Delaware,
Newark, Delaware 19716, USA }

\maketitle

\begin{abstract} A new approach to large-scale nuclear structure
calculations, based on the Density Matrix Renormalization Group
(DMRG), is described. The method is tested in the context of a
problem involving many identical nucleons constrained to move in a
single large-j shell and interacting via a pairing plus quadrupole
interaction. In cases in which exact diagonalization of the
hamiltonian is possible, the method is able to reproduce the exact
results for the ground state energy and the energies of low-lying
excited states with extreme precision. Results are also presented for
a model problem in which exact solution is not feasible.

\end{abstract}

\vspace{0.2in}
%\narrowtext

\begin{center}
\end{center}

\vspace{0.5in}
\begin{multicols}{2}

The nuclear shell model\cite{R1} is arguably the most powerful
approach for a microscopic description of nuclear properties.  In
this approach, the low-energy structure of a given nucleus is
described by assuming an inert doubly-magic core and then seeing how
the effective interaction  scatters the remaining nucleons over the
valence orbits of the next major shell(s). Despite the enormous
simplification provided by this shell-model approach, it is still
only possible to describe nuclei in this way within limited regions
of the periodic table, namely where the number of active nucleons or
the degeneracy of the valence shells is sufficiently small. The most
ambitious implementation of this method to date has been in a
treatment of the binding energies of nuclei in the $fp$ shell through
$^{64}Zn$\cite{R2}.

If we wish to use the shell-model approach in the description of
heavier nuclei or nuclei further from closed shells, we must come up
with a reliable truncation procedure, one that can reduce the number
of shell-model configurations while maintaining the key dynamics of
the interacting nucleons. Historically, many approaches have been
used. Some truncate on the basis of weak-coupling
considerations\cite{R3}, others on the basis of symmetry
considerations\cite{R4}, and others on the basis of Monte Carlo
sampling\cite{R5}. Within the latter approach, it has recently proven
possible to go beyond the $fp$ shell to describe the transition from
spherical to deformed nuclei in the Barium isotopes\cite{R6}. In this
work, we describe an alternative approach to large-scale nuclear
structure calculations, called the Density Matrix Renormalization
Group (DMRG). This method, developed originally in the framework of
low-dimensional quantum lattice systems\cite{R7}, was recently
extended to finite Fermi systems\cite{R8}. The new methodology was
first used in the treatment of a pairing problem\cite{R8} of
relevance to the physics of ultrasmall superconducting grains. Here
we present its first application to a problem of relevance to
nuclear structure.

The basic idea of the DMRG method, as appropriate to finite Fermi
systems, is to {\it systematically} take into account the physics of
{\it all} single-particle levels. This is done by first taking into
account the most important levels, namely those that are nearest to
the Fermi surface, and then gradually including the others in
subsequent iterations. At each step of the procedure, truncation is
implemented so as to optimally take into account the effects of the
levels that are added while keeping the problem tractable.

We will assume from the outset that each single-particle level in the
problem admits the same number of possible states. This is the case,
for example, when working in a single-particle basis of
axially-symmetric Nilsson-like levels, or if we consider pairs of
time reversal states in a spherical basis. Each such level can
accomodate four possible states, one with no particles, two with one
particle, and one with two particles. We denote the number of states
that a given single-particle level admits as {\it s}; in this case,
$s=4$.

Next, we assume that we have already treated some of the levels for
particles and the same number for holes, namely those closest to the
Fermi energy, and that the number of states in the two spaces is the
same.  We call that common number of states $m$. When we add the next
particle level and the next hole level, the number of particle states
increases from $m$ to $s \times m$ and the number of hole states
likewise increases from $m$ to $s \times m$. The DMRG method
truncates from the $s \times m$ states to the {\it optimum} $m$
states, both for particles and for holes.

Following this {\it optimal} truncation, we then add the next levels
for particles and holes and truncate again to the optimum $m$ states
for each. This procedure is continued until all particle and hole
levels have been sampled. [Note: If one type of level is exhausted
before the other, we subsequently add those levels that remain one at
a time.]

Finally, we carry out the calculation as a function of $m$, the
number of particle and hole states that are kept. All previous 
applications of the
DMRG method\cite{R9} have exhibited exponential convergence of the results
(e.g., for the ground state energy) as a function of $m$, suggesting that this 
should likewise be the case in applications to nuclear structure. If so, we stop the
procedure once the changes that arise with increasing $m$ are
acceptably small.

The key question not yet addressed is ``What do we mean by optimum
and how do we implement a truncation to those optimum states?"

To answer these questions, we now consider the ground state of the
full system, expressed as a sum of terms involving states in the
particle space, $|i>_P$, coupled to states in the hole space, $|j>_H
~$,

\begin{displaymath}
|\Psi> ~=~ \sum_{i=1,N_P} \sum_{j=1,N_H} \Psi_{ij} |i>_P |j>_H ~,
\end{displaymath}
where $N_P$ is dimension of the particle subspace and $N_H$ is the
dimension of the hole subspace. [In the previous discussion,
$N_P=N_H=s \times m $.]

What we would like to do is to construct the optimal approximation to
the ground state wave function $|\Psi>$ that is achieved when we only
retain $m$ states in the particle space and $m$ states in the hole
space. By optimal, we will mean that the projected wave function, the
one that arises following the truncation, has the {\it largest
possible overlap} with the exact ground state wave function $|\Psi>$.

We will implement the truncation in two steps, first asking what is
the best approximation when we truncate the particle states and then
what is the best approximation when we truncate the hole space.

To arrive at the optimum truncation for particles, we first introduce
the corresponding ground state reduced density matrix,

\begin{equation}
\rho^P_{ii'}= \sum_{j=1,N_H} \Psi_{ij}  \Psi^{*}_{i'j} ~,
\label{rop}
\end{equation}
obtained by contracting over all the states of the hole space. We
then diagonalize this $N_P \times N_P$ matrix,

\begin{equation}
\rho^P | u^{\alpha} >_P ~=~ \omega^P_{\alpha} ~ |u^{\alpha} >_P ~.
\end{equation}
A given eigenvalue $\omega^P_{\alpha}$ represents the probability of
finding the particle state $|u^{\alpha} >_P$ in the full ground state
wave function of the system. The optimal truncation corresponds to
retaining the $m$ eigenvectors that have the largest probability of
being present in the ground state wave function, or equivalently
those that correspond to the largest eigenvalues $\omega^P_{\alpha}$
\cite{R9}.

Analogously, we construct the ground state reduced density matrix for
holes, 
\begin{equation} \rho^H_{jj'}= \sum_{i=1,N_P} \Psi_{ij}
\Psi^{*}_{ij'} ~, 
\label{roh} 
\end{equation} 
by contracting over
particle states. If we diagonalize the $N_H \times N_H$ density
matrix for holes and retain only the $m$ states with the largest
eigenvalues, we are guaranteed to be choosing the best hole
truncation in the sense of maximal overlap with the exact ground
state.

Summarizing, in each DMRG iteration we add to the system a new
particle level and a new hole level. We then construct the
hamiltonian matrix for the enlarged system and diagonalize it for the
ground state and some low-lying excited states . From the
ground-state wavefuction, we calculate the reduced density matrices
for particles (\ref{rop}) and holes (\ref{roh}) and diagonalize them.
For each, we retain the $m$ eigenvectors corresponding to the $m$
largest eigenvalues. We then transform all operator matrices in the
enlarged particle and hole spaces to the new truncated basis,
completing the iteration.

As noted earlier, the first application of this methodology was
reported in Ref.~\cite{R6} in the context of a pairing hamiltonian
acting over a very large number of equally separated
doubly-degenerate single-particle levels. The hamiltonian for this
so-called picket fence model is

\begin{equation}
H~=~ \sum_{j,\sigma=+,-} ~ \epsilon_{j\sigma} c^{\dagger}_{j\sigma}
~c_{j\sigma} ~-~ \lambda ~ d ~ \sum_{j,j'} c^{\dagger}_{j+}
~c^{\dagger}_{j-}~c_{j'-} ~c_{j'+} ~, \label{hamil}
\end{equation}
with $\epsilon_{j\sigma}~=~ j~d $ and $j~=~ 1,...,\Omega ~.$ Here,
$\Omega$ is the total number of levels and $d$ is the spacing between
adjacent levels The calculations of Ref.~\cite{R6} assumed half
filling, so that the total number of particles distributed over the
$\Omega$ levels is also $\Omega$.

As an example of the quality of the results that can be obtained with
this method, consider the case of $\Omega=400$. For this value of
$\Omega$, the dimension of the full hamiltonian matrix that would
have to be diagonalized is $10^{119}~$! While this problem is
obviously much too large to be solved by standard diagonalization
techniques, it can be solved to arbitrary accuracy using a method
developed by Richardson in the 1960s\cite{R10}. For a problem in
which $\lambda=0.224$, the Richardson solution for the ground state
has a correlation energy of -22.5183141 in units of $d$. When the
same problem is solved using the DMRG method with $m=60$, a ground
state correlation energy of -22.5168 in the same units is achieved.
The agreement is to better than 1 part in $10^4$, despite the fact
that the maximum dimension hamiltonian matrix that had to be treated
for this value of $m$ was only 3066.

Clearly, for a pairing hamiltonian acting in a uniform
doubly-degenerate single-particle space, the Density Matrix
Renormalization Group method works remarkably well. On the other
hand, the hamiltonian (\ref{hamil}) of this problem is extremely
simple, being equivalent to a one-body hamiltonian for a hard-core
boson. In that respect, even though the previous results are suggestive
that the method might also work well for problems in nuclear
structure physics, we still need to demonstrate this more
convincingly by applying it to a fermion problem with a true
two-body interaction.

With that in mind, we have now completed the first test of the new
DMRG methodology in nuclear structure. We considered a schematic
model in which a large number of identical particles are restricted
to a single large-j shell and interact via a sum of a pairing plus
quadrupole force.  Since a single j shell does not give rise
naturally to a Fermi surface, we also included a single-particle
energy term in the hamiltonian, one that favors an oblate solution.
The hamiltonian of the model is
\begin{equation}
 H= - \chi Q \cdot Q - g  P^{\dagger}~ P -
\epsilon \sum_{m} |m| ~ c^\dagger_{jm} c_{jm} ~.
\end{equation}
Because of the last term, the hamiltonian is not rotationally
invariant, so that its eigenstates do not have good angular momentum.

The first results we present are for 10 particles in a $j=25/2$
orbit. The dimension of the hamiltonian matrix that would have to be
diagonalized (in the m scheme) for this problem is 109,583, which can
be readily handled using the Lanczos algorithm.

In Table I, we present the results for $\chi=1,~g=0$, and
$\epsilon=0.1$. The first row gives the exact ground state energy; 
subsequent rows give the ground state energy obtained using the DMRG
approach as a function of $m$. At the end of each row, we show the
maximum dimension hamiltonian matrix that must be diagonalized in the
DMRG procedure.

As is evident from the table, the results converge very rapidly to
the exact ground state energy. By $m$=60, we obtained a result that
is off by only 1 part in $10^6$.   For this choice of $m$, the
largest matrix we had to diagonalize was $398 \times 398$.

To see what happens in the presence of both quadrupole and pairing
correlations, we next present the results for $\chi=1,~ g=0.05$, and
$\epsilon=0.1$. Such a value for the
pairing strength leads to a strong depletion of probability from the
hole levels, but still leaves a well defined Fermi surface.

Results for the ground state energy of the system are summarized in
Table II.  The exact result is $-16.00367$, obtained using
the Lanczos algorithm. In the DMRG approach, we get an energy of
$-16.00151$ for m=80, with a maximum hamiltonian dimension  of 685. The
results continue to improve slightly with increasing $m$, as the
maximum dimension of the hamiltonian  matrix increases.

The DMRG results for the ground state energy are plotted 
in Fig. 1 as a function of
$m$.  Included in the figure is an exponential fit to these results.
When extrapolated, this exponential predicts an asymptotic end result
of $-16.00358 \pm .00027$, in excellent agreement with the exact
ground state energy.

Table III shows results for the excitation energies of the lowest
three excited states. The agreement for the excited states is almost
as good as for the ground state, even though the method, as described
earlier, only targeted the ground state in the optimization
procedure.

The bottom line is that for these calculations, in which we could
compare with the results of Lanczos diagonalization, we obtain
excellent agreement with the exact results, not just for the ground
state but for higher excited states as well. Furthermore, the
excellent results are achieved while diagonalizing matrices of
moderate dimensions.

The fact that we could achieve a high level of accuracy not just for
the ground state but for low-lying excited states as well may be a
reflection that all these states have the same intrinsic structure.
When different intrinsic structures enter, it may be necessary to
modify our optimization criterion to include mixed density matrices
that contain information on more than one state of the
system\cite{R9}.

The excellent quality of the results we obtained for a $j=25/2$ orbit
encouraged us to treat a more complex system, one for which exact
diagonalization is not possible. We considered the case of a $j=55/2$
orbit with 20 particles. The other parameters of the calculation were
$\chi=1,~g=0.1$, and $\epsilon=0.1$.  In this case, the exact
calculation would involve a matrix of dimension 5.31064 $\times$
10$^{13}$.   The results for the ground state energy are shown in
Table IV. By $m=60$, the calculations have clearly converged and we
should have a reliable ground state energy to about six significant
figures. 

In our view, these calculations demonstrate very clearly the great
promise of the DMRG method in nuclear structure. As long as there is
a well defined Fermi surface in the problem, the method leads to
extremely accurate results not only for the ground state but for
low-lying excited states as well. Furthermore, the fairly rapid
convergence that typically arises as a function of $m$ suggests that
the method can be used quite reliably for very large-scale
calculations.

It is worthwhile here to expand briefly on the rationale for the
assumption that all levels of the single-particle basis  admit the
same number of states. This assumption greatly facilitates
implementation of an iterative scheme in which the addition of new
levels can be readily accomodated with no change of formalism.
Relaxation of this assumption may be feasible, but no doubt at a
significant cost to computational simplicity.

Now that we have completed this first test of the methodology with
such impressive success, we are planning to gradually expand the
complexity of the problems we consider. Our ultimate goal of course
is to use this method to treat very large-scale nuclear structure
problems involving both neutrons and protons populating a set of
non-degenerate single-particle levels and interacting via a general
interaction.

This work was supported in part by the National Science Foundation
under grant \# PHY-9970749, by the Spanish DGI under grant
BFM2000-1320-C02-02, and by NATO under grant PST.CLG.977000. One of
the authors (SP) wishes to thank the Institute for Nuclear Theory at
the University of Washington for its hospitality and the Department
of Energy for partial support during the completion of this work.

\end{multicols}

\newpage

\begin{table}[tbp]
\begin{center}
{{\bf Table 1:} Ground state energy for 10 particles in a j=25/2 level. \\
The Hamiltonian parameters are: $x=1$, $g=0$,and  $\epsilon=0.1$.}
\end{center}
\begin{tabular}{lccc}
& $m$ & $E_{gs}$ & $Max(Dim)$ \\
\tableline
& $Exact$ & -15.58837 & 109,583 \\
\hline
& 20 & -15.58798& 106 \\
& 40& -15.58830 & 217 \\
& 60 & -15.58836 & 398 \\
& 80 & -15.58836 & 656 \\
\end{tabular}
\end{table}

\vspace{0.25in}
\begin{table}[tbp]
\begin{center}
{{\bf Table 2:} Ground state energy for 10 particles in a j=25/2 level. \\
The Hamiltonian parameters are: $x=1$, $g=0.05$, $\epsilon=.1$.}
\end{center}
\vspace{0.1in}
\begin{tabular}{lccc}
& $m$ & $E_{gs}$ & $Max(Dim)$ \\
\tableline
& $Exact$ & -16.00367 & 109,583 \\
\hline
& 20 & -15.99380& 118 \\
& 40& -15.99776 & 238 \\
& 60 & -15.99946 & 548 \\
& 80 & -16.00151 & 685 \\
& 100 & -16.00246  & 964 \\
& 120 &-16.00271 & 1267 \\
& 140 &-16.00306 & 1648 \\
\end{tabular}
\end{table}

\vspace{0.25in}

\begin{table}[tbp]
\begin{center}
{{\bf Table 3:} Excitation energies for 10 particles in a $j=25/2$ level.
The hamiltonian parameters are: $x=1$, $g=0.05$ and $\epsilon=0.1$.}
\end{center}
\vspace{0.1in}
\begin{tabular}{lcccc}
& $m$ & $E_1$ & $E_2$ & $E_3$\\
\tableline
& $Exact$ & 0.51643 & 0.87141 &1.10294\\
\hline
& 40& 0.52464 & 0.96894  & 1.18556\\
& 60 & 0.51774  & 0.90385& 1.11528\\
& 80 &0.51854  & 0.88581& 1.11353\\
& 100 & 0.51817  & 0.88274& 1.11432 \\
& 120 & 0.51772 & 0.88068 & 1.11115\\
& 140 & 0.51743 & 0.87842  & 1.10942\\
\end{tabular}
\end{table}

\vspace{0.25in}
\begin{table}[tbp]
\begin{center}
{{\bf Table 4:} Ground state energy for 20 particles in a j=55/2 level. \\
The Hamiltonian parameters are: $x=1$, $g=0.1$, $\epsilon=.1$.}
\end{center}
\vspace{0.1in}
\begin{tabular}{lccc}
& $m$ & $E_{gs}$ & $Max(Dim)$ \\
\tableline
& $Exact$ & ? & 5.31$\times$ 10$^{13}$ \\
\hline
& 20 & -103.98844& 100 \\
& 30& -103.99420& 180 \\
& 40 & -103.99574 & 240 \\
& 50 & -103.99827 & 361 \\
& 60 & -103.99894 & 430 \\
\end{tabular}
\end{table}

\newpage

\begin{figure}
\end{figure}

%\onecolumn
\begin{figure}[t]
\begin{center} \leavevmode
\epsfxsize=15.0cm \epsfbox{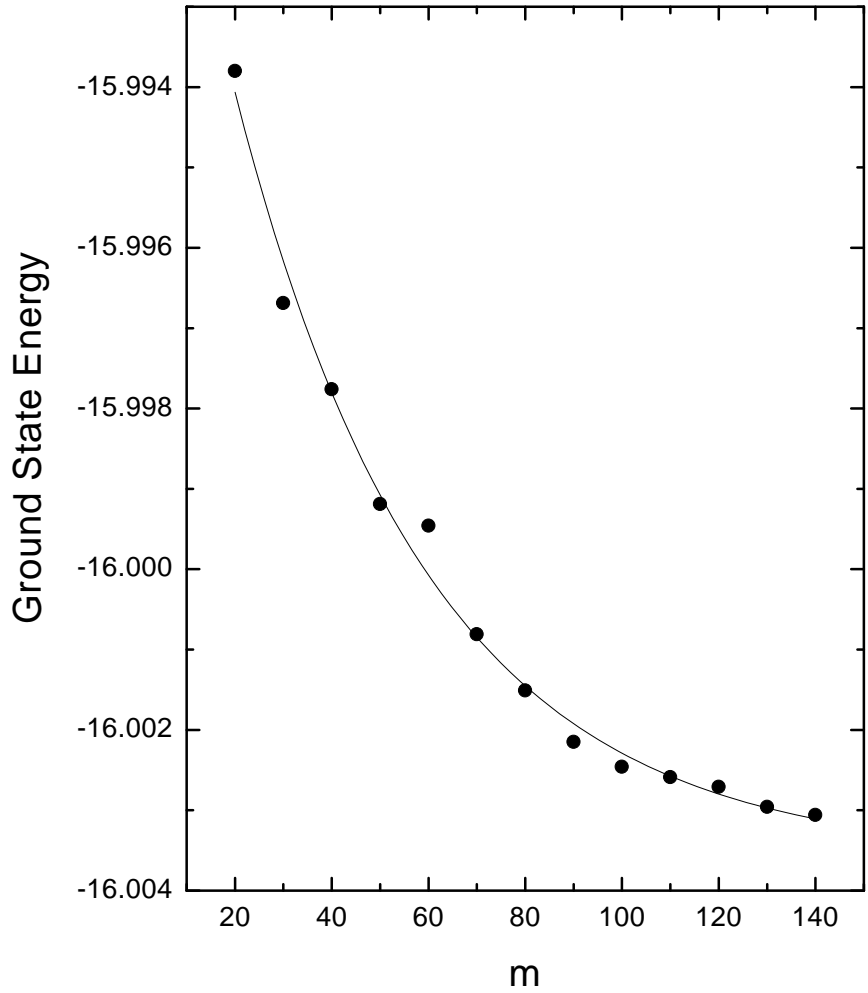}
\caption{Calculated results for the ground state energy
(circles) and an exponential fit to those results (solid curve) for a
system of 10 identical nucleons occupying a $j=25/2$ orbit and
interacting via a hamiltonian with parameters $\chi=1,~g=0.05$ and
$\epsilon=0.1$. }
\end{center}
%\vspace{-0.5cm}
\label{fig:fig1}
\end{figure}

%\end{multicols}

\end{document}